\documentclass[aps,prd,superscriptaddress,showpacs,preprint,amsmath,amssymb]{revtex4}
\usepackage{graphicx, bm}
\usepackage[usenames]{color}

\begin{document}

\draft
\title{Projections for model-independent sensitivity estimates on top-quark anomalous electromagnetic couplings at the FCC-he}

\author{A. A. Billur\footnote{abillur@cumhuriyet.edu.tr}}
\affiliation{\small Deparment of Physics, Sivas Cumhuriyet University, 58140, Sivas, Turkey.\\}

\author{M. K\"{o}ksal\footnote{mkoksal@cumhuriyet.edu.tr}}
\affiliation{\small Deparment of Optical Engineering, Sivas Cumhuriyet University, 58140, Sivas, Turkey.\\}

\author{ A. Guti\'errez-Rodr\'{\i}guez\footnote{alexgu@fisica.uaz.edu.mx}}
\affiliation{\small Facultad de F\'{\i}sica, Universidad Aut\'onoma de Zacatecas\\
         Apartado Postal C-580, 98060 Zacatecas, M\'exico.\\}

\author{ M. A. Hern\'andez-Ru\'{\i}z\footnote{mahernan@uaz.edu.mx}}
\affiliation{\small Unidad Acad\'emica de Ciencias Qu\'{\i}micas, Universidad Aut\'onoma de Zacatecas\\
         Apartado Postal C-585, 98060 Zacatecas, M\'exico.\\}

\date{\today}

\begin{abstract}

The measurement of the top-quark anomalous electromagnetic couplings is one of the most important goals of the top-quark
physics program in present and future collider experiments and would provide direct information on the non-standard
interactions of the top-quark. We study a top-quark pair production scenario at the Future Circular Collider Hadron-Electron (FCC-he) through $e^-p \to e^-\gamma^*\gamma^*p \to e^-t\bar t p$ collisions, which will provide information
on the sensitivities of anomalous $\hat a_V$ and $\hat a_A$ couplings at $95\%\hspace{0.8mm}C.L.$, as well as create the possibility of probing new physics. Energy of the $e^-$ beams is taken to be $E_e=250\hspace{0.8mm}GeV$ and $500\hspace{0.8mm}GeV$, and the energy of the $p$ beams is considered to be $E_p=50\hspace{0.8mm}TeV$. With these energies the FCC-he can measure the dipole moments of the top-quark $\hat a_V$ and $\hat a_A$ with a sensitivity of the order ${\cal O}(10^{-2}-10^{-1})$.\\

\end{abstract}

\pacs{14.65.Ha, 13.40.Em\\
Keywords: Top quarks, Electric and Magnetic Moments.}

\vspace{5mm}

\maketitle

\section{Introduction}

The detection of the Higgs boson \cite{Tanabashi,Englert,Higgs,Higgs1,Guralnik} by the ATLAS \cite{Atlas} and CMS \cite{CMS}
Collaborations at the Large Hadron Collider (LHC), together with the absence so far of any signature new physics beyond the
Standard Model (BSM) in collisions at center-of-mass energies of several $TeV$, has triggered great interest of the scientific
community in the planning of future colliders that increase the energy and precision frontiers in cleaner environments. The Future
Circular Collider (FCC) study develops options for potential high-energy frontier circular colliders at CERN for the post LHC era
that will open up new horizons in the field of fundamental physics. These colliders, with their high precision and high-energy reach,
could extend the search for new particles and interactions well beyond the LHC that may hold the key to understanding and responding
to the open problems of the SM such as: evidence for dark matter, prevalence of matter over antimatter, and neutrino masses. The FCC study
puts great emphasis on the scenarios of high-intensity and high-energy frontier colliders: $pp$, $e^+e^-$ and $e^-p$. It should be mentioned
that in comparison with the LHC, the FCC-he has the advantage of providing a clean environment with small background contributions
from QCD strong interactions. In addition, as the initial states are asymmetric, the backward and forward scattering can be disentangled.

In this paper we considered the electron-proton collision of the FCC-he that is proposed to build on the same site with
LHC, as the future extension of the Large Hadron Electron Collider (LHeC). In FCC-he, construction of an Energy Recovery Linac
(ERL) is proposed to deliver electrons with energies ranging from $E_e=250\hspace{0.8mm}{\rm GeV}$ to $E_e=500\hspace{0.8mm}{\rm GeV}$,
while a proton beam is provided by the main FCC ring  based on a new $80-100\hspace{0.8mm}{\rm km}$ circumference tunnel infrastructure.
The FCC-eh will collide a $250\hspace{0.8mm}{\rm GeV}$ to $500\hspace{0.8mm}{\rm GeV}$ electron beam from a linear accelerator external
and tangential to the main FCC tunnel, with a $50\hspace{0.8mm}{\rm TeV}$ proton beam. In addition, it would collect factors of thousands
more luminosity than the first colliders electron-proton such as the Hadron-Electron Ring Accelerator (HERA). The machine would serve as
the most powerful, high-resolution microscope onto the substructure of matter ever built. High-energy $ep$ collisions would provide precise
information on the quark and gluon structure of the proton, and how they interact. This machine would complement and enhance the study of the
physics of the Higgs boson, top-quark, tau-lepton  and broaden the new physics searches also performed at the Future Circular Collider
Hadron-Hadron (FCC-hh) and the Future Circular Collider Electro-Electron (FCC-ee). Discoveries such as quark substructure might also arise.
In $ep$ collisions new particles can be created in the annihilation of the electron and a (anti)quark, or may be radiated in the exchange of
a photon or other vector bosons. FCC-eh could also provide access to Higgs self-interactions and extended Higgs sectors, including scenarios
involving dark matter. If neutrino oscillations arise from the existence of heavy sterile neutrinos, direct searches at the FCC-eh would have
great discovery prospects in kinematic regions complementary to FCC-hh and FCC-ee, giving the FCC complex a striking potential to shine light
on the origin of neutrino masses. For a complete and detailed study on the physics and detector design concepts see Refs. \cite{FCChe,Fernandez,Fernandez1,Fernandez2,Huan,Acar}.

The aim of this study is to obtain model-independent sensitivity estimates on top-quark anomalous electromagnetic couplings
$\tilde a_V$ and $\tilde a_A$ at the Future Circular Collider Hadron-Electron (FCC-he) \cite{FCChe,Fernandez,Fernandez1,Fernandez2,Huan,Acar,Bruening}
through $e^-p \to e^-\gamma^*\gamma^*p \to e^-t\bar t p$ collisions. We have based our study on the FCC-he which has been designed to
collide electrons with an energy $E_e=250\hspace{0.8mm}GeV$ to $E_e=500\hspace{0.8mm}GeV$ with protons with an energy $E_p=50\hspace{0.8mm}TeV$,
corresponding to the center-of-mass energies $\sqrt{s}=2\sqrt{E_e E_p}= 7.07\hspace{0.8mm}TeV$ and $10\hspace{0.8mm}TeV$. Depending on the energy of the incoming electrons and the design of the collider, the anticipated integrated
luminosity is approximately $50-1000\hspace{0.8mm}fb^{-1}$.

Many studies on the physical potential of top-quark using different processes and environments have been presented by several theoretical,
experimental and phenomenological authors and groups. Furthermore, the processes involving top-quarks provide us unique opportunities to
test the Standard Model (SM) predictions and look for possible signatures of new physics BSM. These signals can be the anomalous
electromagnetic dipole moments of the top-quark, that is, its Magnetic Moment ($t$-MM) and its Electric Dipole Moment ($t$-EDM),
with the latter considered as a source of CP violation. The estimate in the SM for the $t$-MM $(a_t)$ \cite{Benreuther}
and $t$-EDM $(d_t)$ \cite{Hoogeveen,Pospelov,Soni} is given by:

\begin{equation}
\mbox{SM}:
\begin{array}{ll}
a_t= 0.02,  \\
d_t < 10^{-30} (e cm),
\end{array}
\end{equation}

\noindent and the $t$-MM can be tested in the LHC and future colliders such as the Compact Linear Collider (CLIC),
the Large Hadron-Electron Collider (LHeC) and the FCC-he. As shown in Eq. (1), the $t$-EDM value is strongly suppressed
and difficult to measure. What's more, if its value were zero, it wouldn't be possible to measure it at all. However, the $t$-EDM  is very useful for probing new physics BSM.

A summary of sensitivities achievable on the electromagnetic dipole moments of the top-quark is given in Table I.
See Refs. \cite{Ibrahim,Atwood,Polouse,Choi,Polouse1,Aguilar0,Amjad,Juste,Asner,Abe,Aarons,Brau,Baer,Grzadkowski:2005ye}
for other results on the $t$-MM and the $t$-EDM in different contexts.

\begin{table}[!ht]
\caption{Summary of sensitivities achievable on the electromagnetic dipole moments of the top-quark.}
\begin{center}
\begin{tabular}{|c| c| c| c|}
\hline
\hline
{\bf Model}  &    {\bf Theoretical sensitivity: $\hat a_V$, $\hat a_A$}        & {\bf C. L.}  &  {\bf Reference}\\
\hline
\hline
Top-quark pair production at LHC                              &   $ (-0.041, 0.043), (-0.035, 0.038) $ & $68 \%$  & \cite{Juste} \\
\hline
$t\bar t\gamma$ production at LHC                             &   $ (-0.2, 0.2), (-0.1, 0.1) $   & $90 \%$ & \cite{Baur} \\
\hline
Radiative $b\to s\gamma$ transitions at Tevatron and LHC      &   $ (-2, 0.3), (-0.5, 1.5) $   & $90 \%$  & \cite{Bouzas} \\
\hline
Process $pp \to p\gamma^*\gamma^*p\to pt\bar t p $ at LHC     &   $ (-0.6389, 0.0233), (-0.1158, 0.1158) $  & $68 \%$  & \cite{Sh} \\
\hline
Measurements of $\gamma p \to t\bar t$ at LHeC               &   $ (-0.05, 0.05), (-0.20, 0.20)    $ & $90 \%$  & \cite{Bouzas1} \\
\hline
Top-quark pair production $e^+e^- \to t\bar t$  at ILC        &   $ (-0.002, 0.002), (-0.001, 0.001) $  & $68 \%$  & \cite{Aguilar} \\
\hline
Process $\gamma e^- \to \bar t b\nu_e$ at CLIC                &   $ (-0.0258, 0.0350 ), (-0.0301, 0.0301) $  & $95 \%$  & \cite{murat} \\
\hline
Process $e^+ e^- \to e^+\gamma^*e^- \to  \bar t b\nu_e e^+$  at CLIC  &   $ (-0.0609, 0.1081 ), (-0.0777, 0.0777) $   & $95 \%$ & \cite{murat} \\
\hline
Mode $\gamma\gamma \to t\bar t$ at CLIC      &   $ (-0.02203, 0.0020 ), (-0.0206, 0.0206)  $   & $95 \%$  & \cite{Billur} \\
\hline
Mode $e^+\gamma \to e^+\gamma^* \gamma \to e^+ t \bar t$ at CLIC    &   $ (-0.4570, 0.0045 ), (-0.0431, 0.0431)  $   & $95 \%$  & \cite{Billur} \\
\hline
Mode $e^+e^- \to e^+\gamma^* \gamma^* e^- \to e^+ t \bar t e^-$ at CLIC &     $ (-0.6013, 0.0151 ), (-0.0890, 0.0890)    $  & $95 \%$  & \cite{Billur} \\
\hline\hline
\end{tabular}
\end{center}
\end{table}

The structure of this paper is as follows. In Section II, we introduce the top-quark effective electromagnetic interactions. In Section III,
we present sensitivity estimates on top-quark anomalous electromagnetic couplings through $e^-p \to e^-\gamma^*\gamma^*p \to e^-t\bar t p$ collisions.
In Section IV, we present our conclusions .

\section{Production of $t\bar t$ pairs via the process $e^-p \to e^-\gamma^*\gamma^*p \to e^-t\bar t p$ }

\subsection{Top-quark Effective Coupling $t\bar t \gamma$}

A suitable model-independent formalism for describing possible new physics effects is based on effective Lagrangian.
In this formalism, all the heavy degrees of freedom are integrated out to obtain the effective interactions between
the SM particles. This is justified due to the fact that the related observables have so far not shown any significant deviation from
the SM predictions. In the effective Lagrangian formalism, potential deviations from the SM for the anomalous $t\bar t\gamma$
coupling could be described using the following Lagrangian:

\begin{equation}
{\cal L}_{eff}={\cal L}_{SM} + \sum_n \frac{\alpha_n}{\Lambda^2}{\cal O}^{(6)}_n + h.c..
\end{equation}

\noindent Here, ${\cal L}_{eff}$ is the effective Lagrangian which contains a series of higher-dimensional operators
built with the SM fields, ${\cal L}_{SM}$ is the renormalizable SM Lagrangian, $\Lambda$ is the scale at which new physics
is expected to be observed, $\alpha_n$ are dimensionless coefficients and ${\cal O}^{(6)}_n$ represents the dimension-six gauge-invariant
operator.

The most general effective coupling $t\bar t\gamma$ includes the SM coupling and contributions from dimension-six effective operators and can be written as \cite{Sh,Kamenik,Baur,Aguilar,Aguilar1}:

\begin{equation}
{\cal L}_{t\bar t\gamma}=-g_eQ_t\bar t \Gamma^\mu_{ t\bar t  \gamma} t A_\mu,
\end{equation}

\noindent where $g_e$ is the electromagnetic coupling constant, and $Q_t$ is the top-quark electric charge.
The Lorentz-invariant vertex function $\Gamma^\mu_{t\bar t \gamma}$, which describes the
interaction of a $\gamma$ photon with two top-quarks, can be parameterized by:

\begin{equation}
\Gamma^\mu_{t\bar t\gamma}= \gamma^\mu + \frac{i}{2m_t}(\hat a_V + i\hat a_A\gamma_5)\sigma^{\mu\nu}q_\nu,
\end{equation}

\noindent where $m_t$ is the mass of the top-quark, $q$ is the momentum transfer to the photon, and the couplings
$\hat a_V$ and $\hat a_A$ are real and related to the anomalous magnetic moment $(a_t)$ and the electric dipole moment
$(d_t)$ of the top-quark, respectively. The relations between $\hat a_V (\hat a_A) $ and $a_t (d_t)$ are given by:

\begin{eqnarray}
\hat a_V&=&Q_t a_t,  \\
\hat a_A&=&\frac{2m_t}{e}d_t.
\end{eqnarray}

The operators that contribute to top-quark eletromagnetic anomalous couplings \cite{Buhmuller,Aguilar2} are:

\begin{eqnarray}
{\cal O}_{uW}^{33}=\bar q_{L3}\sigma^{\mu\nu}\tau^a t_{R}{\tilde \phi} W_{\mu\nu}^{a},
\end{eqnarray}

\begin{eqnarray}
{\cal O}_{uB\phi}^{33}=\bar q_{L3}\sigma^{\mu\nu}t_{R}{\tilde \phi} B_{\mu\nu},
\end{eqnarray}

\noindent where $\bar q_{L3}$ is the quark field, $\sigma^{\mu\nu}$ are the Pauli matrices and ${\tilde \phi}$ is the Higgs doublet,
while $W_{\mu\nu}^{a}$ and $B_{\mu\nu}$ are the $U(1)_Y$ and $SU(2)_L$ gauge field strength tensors, respectively.

From the parametrization given by Eq. (3), and from the operators of dimension-six given in Eqs. (7) and (8) we obtain
the corresponding CP even ${\hat a_V}$ and CP odd ${\hat a_A}$ observables:

\begin{eqnarray}
\hat a_V=\frac{2 m_t}{e} \frac{\sqrt{2}\upsilon}{\Lambda^{2}} Re\Bigl[\cos\theta _{W} C_{uB\phi}^{33} + \sin\theta _{W} C_{uW}^{33}\Bigr],
\end{eqnarray}

\begin{eqnarray}
\hat a_A=\frac{2 m_{\tau}}{e} \frac{\sqrt{2}\upsilon}{\Lambda^{2}} Im\Bigl[\cos\theta _{W} C_{uB\phi}^{33} + \sin\theta _{W} C_{uW}^{33}\Bigr].
\end{eqnarray}

\noindent In these equations, $\upsilon=246$ GeV is the breaking scale of the electroweak symmetry, $\Lambda$ is the new physics scale, and
$\sin\theta _{W} (\cos\theta _{W})$ is the sine(cosine) of the weak mixing angle.

\subsection{Theoretical Calculations}

In the $e^-p$ colliders, the top-quark pairs can be produced through the channel $e^-p \to e^-\gamma^*\gamma^*p \to e^-t\bar t p$. The schematic diagram corresponding to this process is given in Fig. 1. The representative leading order Feynman diagrams for the
subprocess $\gamma^*\gamma^* \to t\bar t$ are depicted in Fig. 2.

For the $\gamma^{*}\gamma^{*}$ collision, including the effects of the effective Lagrangian given by Eq. (3), the corresponding matrix elements for the subprocess $\gamma^*\gamma^* \to t\bar t$ are given as a function of the Mandelstam invariants $\hat{s}$, $\hat{t}$ and $\hat{u}$, as well as of their dipole moments $\hat a_V$ and $\hat a_A$:

\begin{eqnarray}
|M_{1}|^{2}&=&\frac{16\pi^{2}Q_{t}^2\alpha^{2}_e}{2m_{t}^{4}(\hat{t}-m_{t}^{2})^{2}}\biggl[48\hat{a}_{V}(m_{t}^{2}-\hat{t})
(m_{t}^{2}+\hat{s}-\hat{t})m_{t}^{4}-16(3m_{t}^{4}-m_{t}^{2}\hat{s}+\hat{t}(\hat{s}+\hat{t})) m_{t}^{4}\nonumber\\
&+&2(m_{t}^{2}-\hat{t})(\hat{a}_{V}^{2}(17m_{t}^{4}+(22\hat{s}-26\hat{t})m_{t}^{2} +\hat{t}(9\hat{t}-4\hat{s}))  \nonumber\\
&+&\hat{a}_{A}^{2}(17m_{t}^{2}+4\hat{s}-9\hat{t})(m_{t}^{2}-\hat{t}))m_{t}^{2}+12\hat{a}_{V}(\hat{a}_{V}^{2}+\hat{a}_{A}^{2})\hat{s}(m_{t}^{3}-m_{t}\hat{t})^{2}\nonumber\\
&-&(\hat{a}_{V}^{2}+\hat{a}_{A}^{2})^{2}(m_{t}^{2}-\hat{t})^{3}(m_{t}^{2}-\hat{s}-\hat{t})\biggr],
\end{eqnarray}

\begin{eqnarray}
|M_{2}|^{2}&=&\frac{-16\pi^{2}Q_{t}^2\alpha^{2}_e}{2m_{t}^{4}(\hat{u}-m_{t}^{2})^{2}}\biggl[48\hat{a}_{V}(m_{t}^{4}+(\hat{s}-2\hat{t})m_{t}^{2}+\hat{t}(\hat{s}+\hat{t}))m_{t}^{4}\nonumber\\
&+&16(7m_{t}^{4}-(3\hat{s}+4\hat{t})m_{t}^{2}+\hat{t}(\hat{s}+\hat{t})) m_{t}^{4}\nonumber\\
&+&2(m_{t}^{2}-\hat{t})(\hat{a}_{V}^{2}(m_{t}^{4}+(17\hat{s}-10\hat{t})m_{t}^{2}+9\hat{t}(\hat{s}+\hat{t})) \nonumber\\
&+&\hat{a}_{A}^{2}(m_{t}^{2}-9\hat{t})(m_{t}^{2}-\hat{t}-\hat{s}))m_{t}^{2}\nonumber\\
&+&(\hat{a}_{V}^{2}+\hat{a}_{A}^{2})^{2}(m_{t}^{2}-\hat{t})^{3}(m_{t}^{2}-\hat{s}-\hat{t})\biggr],
\end{eqnarray}

\begin{eqnarray}
M_{1}^{\dag}M_{2}+M_{2}^{\dag}M_{1}&=&\frac{16\pi^{2}Q_{t}^2\alpha^{2}_e}{m_{t}^{2}(\hat{t}-m_{t}^{2})(\hat{u}-m_{t}^{2})} \nonumber \\
&\times &\biggl[-16(4m_{t}^{6}-m_{t}^{4}\hat{s})+8\hat{a}_{V}m_{t}^{2}(6m_{t}^{4}-6m_{t}^{2}(\hat{s}+2\hat{t})-\hat{s})^{2} \nonumber \\ &+&6\hat{t})^{2}+6\hat{s}\hat{t})+(\hat{a}_{V}^{2}(16m_{t}^{6}-m_{t}^{4}(15\hat{s}+32\hat{t})+m_{t}^{2}(15\hat{s})^{2} \nonumber \\
&+&14\hat{t}\hat{s}+16\hat{t})^{2})+\hat{s}\hat{t}(\hat{s}+\hat{t}))+\hat{a}_{A}^{2}(16m_{t}^{6}-m_{t}^{4}(15\hat{s}+32\hat{t})  \nonumber\\
&+&m_{t}^{2}(5\hat{s})^{2}+14\hat{t}\hat{s}+16\hat{t})^{2})+\hat{s}\hat{t}(\hat{s}+\hat{t})))-4\hat{a}_{V}\hat{s}(\hat{a}_{V}^{2}+\hat{a}_{A}^{2})\nonumber\\
&\times& (m_{t}^{4}+m_{t}^{2}(\hat{s}-2\hat{t})+\hat{t}(\hat{s}+\hat{t}))-4\hat{a}_{A}(\hat{a}_{V}^{2}+\hat{a}_{A}^{2})(2m_{t}^{2}
-\hat{s}-2\hat{t}) \nonumber\\
&\times& \epsilon_{\alpha \beta \gamma \delta} p_{1}^{\alpha}p_{2}^{\beta}p_{3}^{\gamma}p_{4}^{\delta}-2\hat{s}(\hat{a}_{V}^{2}+\hat{a}_{A}^{2})^{2}
(m_{t}^{4}-2\hat{t}m_{t}^{2}+\hat{t}(\hat{s}+\hat{t}))\biggr].
\end{eqnarray}

\noindent In Eqs. (11)-(13), $\hat s=(p_1 + p_2)^2=(p_3 + p_4)^2$, $\hat t=(p_1 - p_3)^2=(p_4 - p_2)^2$, $\hat u=(p_3 - p_2)^2=(p_1 - p_4)^2$,
where $p_{1}$ and $p_{2}$ are the four-momenta of the incoming photons, $p_{3}$ and $p_{4}$ are the momenta of the outgoing top-quarks,
$Q_{t}$ is the top-quark charge, $\alpha_e=g^2_e/4\pi$ is the fine-structure constant and $m_t$ is the mass of the top-quark.

The Weizsacker-Williams Approximation (WWA), also alternatively called the Equivalent Photon Approximation (EPA) \cite{Budnev,Baur1,Piotrzkowski},
is useful for determining the $e^-p \to e^-\gamma^*\gamma^*p \to e^-t\bar t p$ scattering cross-section. In EPA, photons emitted from incoming
charged particles with very low virtuality are scattered at very small angles from the beam pipe. Because the emitted quasi-real photons have
a low $Q^{2}$ virtuality, they are almost real. These processes have been observed experimentally at the LEP, the Tevatron and the LHC \cite{Abulencia,Aaltonen1,Aaltonen2,Chatrchyan1,Chatrchyan2,Abazov,Chatrchyan3}. In WWA or EPA, the quasi-real photons emitted from both lepton
and proton beams collide with each other and produce the subprocess $\gamma^{*} \gamma^{*} \rightarrow t \bar{t}$. In the WWA, the spectrum of
the photon emitted by electron $(\gamma^*_1)$ is given by \cite{Belyaev,Budnev}:

\begin{eqnarray}
f_{\gamma^{*}_{1}}(x_{1}) &=& \frac{\alpha}{\pi E_{e}}
\left\{ \left[\frac{1-x_{1}+x_{1}^{2}/2}{x_{1}}\right] \log{\left(\frac{Q_{max}^{2}}{Q_{min}^{2}}\right)}
-\frac{m_{e}^{2}x_{1}}{Q_{min}^{2}} \left(1-\frac{Q_{min}^{2}}{Q_{max}^{2}}\right) \right.   \nonumber \\
&-& \left. \frac{1}{x_{1}}\left[1-\frac{x_{1}}{2}\right]^{2}log\left(\frac{x_{1}^{2}E_{e}^{2}
+Q_{max}^{2}}{x_{1}^{2}E_{e}^{2}+Q_{min}^{2}}\right) \right\}, \nonumber \\
\end{eqnarray}

\noindent where $x_1=E_{\gamma^*_1}/E_{e}$ and $Q^2_{max}$ is maximum virtuality of the photon. The minimum value of the $Q^2_{min}$
is given by:

\begin{eqnarray}
Q_{min}^{2}=\frac{m_{e}^{2}x_{1}^{2}}{1-x_{1}}.
\end{eqnarray}

The spectrum of the photon emitted by proton $(\gamma^*_2)$ can be written as follows \cite{Belyaev,Budnev}:

\begin{eqnarray}
f_{\gamma^{*}_{2}}(x_{2})=\frac{\alpha}{\pi E_{p}}\left\{\left[1-x_{2}\right]\left[\varphi\left(\frac{Q_{max}^{2}}{Q_{0}^{2}}\right)-\varphi\left(\frac{Q_{min}^{2}}{Q_{0}^{2}}\right)\right]\right\},
\end{eqnarray}

\noindent where the function $\varphi$ is given by:

\begin{eqnarray}
\varphi(\theta)=&&(1+ay)\left[-\textit{In}\left(1+\frac{1}{\theta}\right)+\sum_{k=1}^{3}\frac{1}{k(1+\theta)^{k}}\right]
+\frac{y(1-b)}{4\theta(1+\theta)^{3}} \nonumber \\
&& +c(1+\frac{y}{4})\left[\textit{In}\left(\frac{1-b+\theta}{1+\theta}\right)+\sum_{k=1}^{3}\frac{b^{k}}{k(1+\theta)^{k}}\right].
\end{eqnarray}

From Eq. (17), we define the following:

\begin{eqnarray}
y=\frac{x_{2}^{2}}{(1-x_{2})},
\end{eqnarray}

\begin{eqnarray}
a=\frac{1+\mu_{p}^{2}}{4}+\frac{4m_{p}^{2}}{Q_{0}^{2}}\approx 7.16,
\end{eqnarray}

\begin{eqnarray}
b=1-\frac{4m_{p}^{2}}{Q_{0}^{2}}\approx -3.96,
\end{eqnarray}

\begin{eqnarray}
c=\frac{\mu_{p}^{2}-1}{b^{4}}\approx 0.028.
\end{eqnarray}

Therefore, the total cross-section of the $e^-p \to e^-\gamma^*\gamma^*p \to e^-t\bar t p$ signal in the WWA is obtained as follows:

\begin{equation}
\sigma(e^{-} p \rightarrow e^{-}\gamma^{*} \gamma^{*} p \rightarrow e^{-} t \bar{t} p)= \int_{x_{1}^{min}}^{x_{1}^{max}} \int_{x_{2}^{min}}^{x_{2}^{max}}f_{\gamma_{1}^{*}}(x_{1}) f_{\gamma_{2}^{*}}(x_{2})d\hat{\sigma}(\gamma^{*}\gamma^{*} \rightarrow t \bar{t}) dx_{1}dx_{2}.
\end{equation}

The main anomalous electromagnetic couplings affecting top-quark physics that are of interest for our study are
$\hat a_V$ and $\hat a_A$. We calculate the dependencies of the $e^-p \to e^-\gamma^*\gamma^*p \to e^-t\bar t p$
associated production cross-sections for FCC-he at $7.07\hspace{0.8mm}TeV$ and $10\hspace{0.8mm}TeV$ on $\hat a_V$ and $\hat a_A$
using CalcHEP \cite{Belyaev}, obtaining the following numerical results: \\

$i)$ For $\sqrt{s}=7.07\hspace{0.8mm} TeV$.

\begin{eqnarray}
\sigma(\hat a_V)&=&\Bigl[(0.00458)\hat a_V^4 + (0.00829)\hat a^3_V + (0.01158)\hat a^2_V +(0.00439)\hat a_V  +0.00165 \Bigr] (pb)   \\
\sigma(\hat a_A)&=&\Bigl[(0.00458)\hat a_A^4 + (0.00937)\hat a^2_A + 0.00165 \Bigr] (pb).
\end{eqnarray}

$ii)$ For $\sqrt{s}=10\hspace{0.8mm} TeV$.

\begin{eqnarray}
\sigma(\hat a_V)&=&\Bigl[(0.00928)\hat a_V^4 + (0.01263)\hat a^3_V + (0.01712)\hat a^2_V +(0.00596)\hat a_V  +0.00225 \Bigr] (pb)   \\
\sigma(\hat a_A)&=&\Bigl[(0.00928)\hat a_A^4 + (0.01414)\hat a^2_A + 0.00225 \Bigr] (pb).
\end{eqnarray}

The sensitivities on the total cross-section and on the coefficients of $\hat a_V$ and $\hat a_A$ increase
with the center-of-mass energy, confirming the expected competitive advantage of the high-energies attainable with the FCC-he.

For signal production, as shown in Figs. 3 and 4, we give the total cross-section of the  process $e^-p \to e^-\gamma^*\gamma^*p \to e^-t\bar t p$
as a function of the anomalous couplings $\hat a_V$ and $\hat a_A$ corresponding to the effective vertex $t\bar t\gamma$. We maintain the energy of the $e^-p$ collider at $\sqrt{s}=7.07\hspace{0.8mm}TeV$ and $\sqrt{s}=10\hspace{0.8mm}TeV$, the two
main options of FCC-he.

As shown in Figs. 3 and 4, the anomalous $\hat a_V$ and $\hat a_A$ parameters have different CP properties which can be
seen in Eqs. (11)-(13) and in Eqs. (23)-(26). The contribution of $\hat a_V$ coupling to the total cross-section is proportional
to even and odd powers. In Fig. 3, there are small intervals around $\hat a_V$ in which the cross-section that includes new physics
is smaller than the SM cross-section (from Eqs. (23) and (25), terms depending on $\hat a_V$ give purely the contribution BSM, and
those which do not dependent on $\hat a_V$ give the SM cross-section). For this reason, the $\hat a_V$ coupling has a partially
destructive effect on the total cross-section. Furthermore, in Fig. 4 the total cross-section with respect to the $\hat a_A$ parameter
is of even power and a nonzero value of this parameter allows a constructive effect on the total cross-section.

In order to illustrate the contribution of the two interactions, in Figs. 5 and 6 we give the total cross-sections with respect
to $\hat a_V$ and $\hat a_A$ for center-of-mass energies $\sqrt{s}=7.07\hspace{0.8mm}TeV$ and $10\hspace{0.8mm}TeV$, respectively.
As expected, the results are similar in both the cases. In Fig. 6, with an increase in $\sqrt{s}$, the value in the total cross-section
rapidly increases and the maximum is achieved either for $\hat a_V = 1$ and $\hat a_A = 1$, or for $\hat a_V = 1$ and $\hat a_A = -1$.
Both figures show great sensitivity with respect to the anomalous couplings $\hat a_V$ and $\hat a_A$.

To put our results in perspective with those reported in the literature, we make a direct comparison of our results for
the total cross-section as a function of the dipole moments $\hat a_V$ and $\hat a_A$ given by Figs. 3 and 4 (or similarly
by Figs. 5 and 6), with those reported in Ref. \cite{Sh} (see Figs. \ref{Fig.3} and \ref{Fig.4}). We find that our results,
using the process $e^-p \to e^-\gamma^*\gamma^*p \to e^-t\bar t p$ at the FCC-he energies, show significant improvement as compared to the process $pp\to p\gamma^* \gamma^* p \to pt\bar t p$ at LHC energies. For example, with the process
that we have considered in this paper, the total cross-section is a factor ${\cal O}(10^3)$ between $pp\to p\gamma^* \gamma^* p \to pt\bar t p$
and $e^-p \to e^-\gamma^*\gamma^*p \to e^-t\bar t p$, that is, our results are 3 orders of magnitude higher than those reported
in Ref. \cite{Sh}. This indicates that the sensitivity on the anomalous couplings $\hat a_V$ and $\hat a_A$ can be improved at the
FCC-he by a few orders of magnitude in comparison with the LHC. In the case of the LHeC, the authors of Ref. \cite{Bouzas1}
specifically measure $\sigma(\gamma e^- \to t\bar t)$ with $10\%$ $(18\%)$ error, obtaining the following results for the $t$-MM $(\kappa)$
and the $t$-EDM $(\tilde\kappa)$: $|\kappa|< 0.05 (0.09)$ and $|\tilde\kappa|< 0.20 (0.28)$. In our case, using the process
$e^-p \to e^-\gamma^*\gamma^*p \to e^-t\bar t p$, we obtain: $\hat a_V=(-0.5014, 0.0457)$ and $\hat a_A=|0.1470|$ with
$\delta_{sys}=5\%$, ${\cal L}=1000\hspace{0.8mm}fb^{-1}$ and $95\%\hspace{0.8mm}C.L.$. Although the conditions for the study
of both processes $\gamma e^- \to t\bar t$ and $e^-p \to e^-\gamma^*\gamma^*p \to e^-t\bar t p$ are not the same, our result
indicates an improvement of the sensitivity by a factor of 0.914 (0.735) for $\hat a_V ( \hat a_A)$ with respect to the results
reported in Ref. \cite{Bouzas1}.

\section{Projections on the dipole moments of the top-quark}

We evaluate the potential of top-quark pairs production at the FCC-he with center-of-mass energies of 7.07 and 10 TeV for the
measurement of the MM and EDM $\hat a_V$ and $\hat a_A$ of the top-quark. To carry out our study, we concentrate on the double top-quark
production process, $e^-p \to e^-\gamma^*\gamma^*p \to e^-t\bar t p$, followed by $t(\bar t)\to bW^+(\bar bW^-)$, where the $W$  boson
decays into leptons and hadrons.

To probe the sensitivity on the anomalous $\hat a_V$ and $\hat a_A$ parameters we perform a $\chi^2$ test and
calculate $68\%$, $90\%$ and $95\%$ confidence level (C.L.) sensitivities. The $\chi^2$ distribution \cite{murat,Billur}
is defined by:

\begin{equation}
\chi^2=\biggl(\frac{\sigma_{SM}-\sigma_{BSM}(\sqrt{s}, \hat a_V, \hat a_A)}{\sigma_{SM}\sqrt{(\delta_{st})^2+(\delta_{sys})^2}}\biggr)^2,
\end{equation}

\noindent with $\sigma_{BSM}(\sqrt{s}, \hat a_V, \hat a_A)$ as the total cross-section which contains contributions from the SM
and physics BSM, $\delta_{st}=\frac{1}{\sqrt{N_{SM}}}$ and $\delta_{sys}$ are the statistical and systematic uncertainties,
respectively. The number of events for the process $e^-p \to e^-\gamma^*\gamma^*p \to e^-t\bar t p$ is given by
$N_{SM}={\cal L}_{int}\times \sigma_{SM} \times BR \times \epsilon_{b-tag}\times \epsilon_{b-tag}$, where ${\cal L}_{int}$ is
the integrated luminosity and  $b$-jet tagging efficiency is $\epsilon_b=0.8$ \cite{atlas}. The top-quark decay almost $100\%$
into a $W$ boson and a $b$ quark. For top-quark pair production, we can categorize decay products according to the decomposition of $W$.
In this work, we assume that for the signal, one of the $W$ bosons decays leptonically and the other hadronically. This event has
already been studied by ATLAS and CMS Collaborations \cite{cmstop1,cmstop2,atlastop}. It is worth mentioning that the branching
rations for $W$ decay are: $BR(W \to q q')=0.674$ for hadronic decay, $BR(W \to l\nu_{e, \mu})=0.213$ for light leptonic decays
and $BR(W \to \tau\nu_\tau)=0.113$ \cite{Data2018}. Therefore, for $t\bar t$ production followed by $t \to Wb$ decays, the branching
rations for the dominant channels are the hadronic and semileptonic, respectively. Thus, we assume that the branching ratio
of the top-quark pair in the final state is $BR(t\to Wb) = 0.286$.

An important part of our study is the incorporation of theoretical uncertainties as there may
be several experimental and systematic uncertainty sources in top-quark identification. In hadron colliders, especially the LHC, the process of determining the cross-section of top pair production has been experimentally studied \cite{topuncertanintyatlas,topuncertaintycms}.
From these studies, the total systematic uncertainty value is about 10$\%$ and is increasingly improved when it is
compared with previous experimental studies \cite{cmstop2}.

Tables II-VII show the projections for model-independent sensitivity on the dipole moments $\hat a_V$ and $\hat a_A$
of the top-quark at the FCC-he. We assume center-of-mass energies of $7.07\hspace{0.8mm}TeV$ and $10\hspace{0.8mm}TeV$,
integrated luminosity ${\cal L}=50, 100, 300, 500, 1000\hspace{0.8mm}fb^{-1}$, systematic uncertainties $\delta_{sys}=0\%, 3\%, 5\%$,
and we determine $68\%$, $90\%$ and $95\%$ C.L. sensitivities. We find that the mode of top-quark pair production
$e^-p \to e^-\gamma^*\gamma^*p \to e^-t\bar t p$ imposes stronger sensitivities on the dipole moments. In conclusion,
the FCC-he can measure the electromagnetic dipole moments of the top-quark with a sensitivity of the order ${\cal O}(10^{-2}-10^{-1})$ at $95\%\hspace{0.8mm}C.L.$.

It is worthwhile to compare the results obtained here with those of Ref. \cite{Sh} which consider the process
$pp\to p\gamma^* \gamma^*  p \to pt\bar t p$ with the LHC running at $\sqrt{s}=14, 33\hspace{0.8mm}TeV$ and with integrated luminosities
of ${\cal L}=100, 300,3000\hspace{0.8mm}fb^{-1}$. The authors of Ref. \cite{Sh} find constraints at $68\%\hspace{0.8mm}C.L.$
of the order ${\cal O}(10^{-2}-10^{-1})$. We also note that, while we do consider three systematic errors in our study, the quoted
sensitivities in Ref. \cite{Sh} do not include theoretical uncertainty. Furthermore, the FCC-he sensitivity is even higher in our process than in that reported in Ref. \cite{Sh}.

Figs. 7 and 8 show the sensitivity contours at the $95\% \hspace{1mm}C.L.$ in the $\hat a_V-\hat a_A$ plane through the process
$e^-p \to e^-\gamma^*\gamma^*p \to e^-t\bar t p$ for $\sqrt{s}=7.07\hspace{0.8mm}TeV,  10\hspace{0.8mm}TeV$ and
${\cal L}=50, 250,1000\hspace{0.8mm}fb^{-1}$ at the FCC-he. We find that the sensitivity of the anomalous couplings can be increased
with an increase in the luminosity as the couplings scales as $1/\sqrt{\cal L}$ for a given $\chi^2$. Thus for $1000 fb^{-1}$,  the
sensitivity of $\hat a_V$ and $\hat a_A$ is improved with respect to $50 fb^{-1}$ and $250 fb^{-1}$, respectively. Concentrating on
the sensitivity at $95\%$ C.L., we observe that only contributions to the dipole moments at the order $\hat a_V=(-0.5900, 0.1188)$
and $\hat a_A=|0.2567|$ could be detected with $50\hspace{0.8mm}fb^{-1}$. With higher luminosity, these sensitivities improve up
to $\hat a_V=(-0.4864, 0.0332)$ and $\hat a_A=|0.1233|$ (at $1000\hspace{0.8mm}fb^{-1}$) (see Table VII). Compared with the sensitivities
of Table I, which arise in different contexts, a significant improvement can be obtained.

We use channel $e^-p \to e^-\gamma^*\gamma^*p \to e^-t\bar t p$ to make fits to estimate the sensitivities at the various
FCC-he center-of-mass energies, using the dependence of the total cross-section on the parameters $\hat a_V $ and $\hat a_A$,
as shown by Eqs. (23)-(26). In making our $\chi^2$ fits, we adopt statistical errors $\delta_{st}=\frac{1}{\sqrt{N_{SM}}}$
(see Eq. (27) ), and we assume systematic uncertainties $\delta_{sys}=0, 3\%, 5\%$. The sensibility results at $68\%$, $90\%$ and $95\%$ C.L.
are plotted in Figs. 9-12 for  $7.07\hspace{0.8mm}TeV$ and $10\hspace{0.8mm}TeV$, respectively.

The results from our $7.07\hspace{0.8mm}TeV$ fit shown in Figs. 9 and 10 include the individual sensitivity obtained considering
just one parameter at a time and the estimated sensitivities are color-coded in red, green and blue. In these figures, the sensitivities
are obtained by taking into account the systematic uncertainties $\delta_{sys} = 0\%, 3\%, 5\%$ at $68\%$, $90\%$ and $95\%$ C.L., respectively.

Similar conclusions hold for the $10\hspace{0.8mm}TeV$ fits whose results are shown in Figs. 11 and 12, where the increases in
sensitivity are even more notable with individual sensitivities on $\hat a_V$ and $\hat a_A$. This level of sensitivity becomes
comparable to that at which future electroweak precision tests may constrain the anomalous couplings $\hat a_V$ and $\hat a_A$
as is the case of the ILC and CLIC \cite{Aguilar,murat,Billur}.

\section{Conclusions}

In this paper, we have emphasized the potential importance for sensitivity of the FCC-he, that is, of a future high-energy $e^-p$ collider
for directly probing possible new physics BSM by studying sensitivity on the total cross-section and on the $\hat a_V$ and $\hat a_A$ at
$\sqrt{s}=7.07, 10\hspace{0.8mm}TeV$ and ${\cal L}=50$ to $1000\hspace{0.8mm}fb^{-1}$. We have stressed and shown numerically
(see Figs. 3-12 and Tables II-VII), that the increase sensitivity at high energy and high luminosity provides opportunities in the process
$e^-p \to e^-\gamma^*\gamma^*p \to e^-t\bar t p$ in particular. The extracted results for the FCC-he, presented in Figs 3-12 and Tables II-VII,
show very high sensitivity, and in some cases improvements are expected with respect to the potential sensitivity for the LHC, ILC and CLIC
(see Table I and Refs. \cite{Ibrahim,Atwood,Polouse,Choi,Polouse1,Aguilar0,Amjad,Juste,Asner,Abe,Aarons,Brau,Baer,Grzadkowski:2005ye}).
Our results motivate more detailed studies including additional benchmark analyses based on full FCC-he detector simulations
at high energies with the aim of verifying and refining our estimates on the sensitivity with which the cross-sections and
the $t$-MM and $t$-EDM for the process $e^-p \to e^-\gamma^*\gamma^*p \to e^-t\bar t p$ could be measured at the FCC-he.
It is worth mentioning that, in comparison with the LHC, the FCC-he has the advantage of providing a clean environment with
small background contributions from QCD strong interactions. With all these arguments already presented, we conclude that the
FCC-he is a viable option for model-independent sensitivity estimates on top-quark anomalous electromagnetic couplings with
very good accuracy.

\begin{table}[!ht]
\caption{Sensitivity on the $\hat a_V$ magnetic moment and the $\hat a_A$ electric dipole moment for the process
$e^-p \to e^-\gamma^*\gamma^*p \to e^-t\bar t p$ for $\sqrt{s}=7.07\hspace{0.8mm}TeV$ and ${\cal L}=50, 100, 300, 500,
1000\hspace{0.8mm}fb^{-1}$ at $68\%$ C.L..}
\begin{center}
 \begin{tabular}{cccc}
\hline\hline
\multicolumn{4}{c}{$\sqrt{s}=7.07\hspace{0.8mm}TeV$,   \hspace{1cm}  $68\%$ C.L.}\\
 \hline
 \cline{1-4} ${\cal L}\hspace{0.8mm}(fb^{-1})$  & \hspace{1.5cm} $\delta_{sys}$ & \hspace{1.5cm} $\hat a_V$   & $\hspace{1.7cm} |\hat a_A|$ \\
\hline
50   &\hspace{1.2cm} $0\%$   &\hspace{1.2cm} [-0.6190, 0.0791]    & \hspace{1.5cm}   0.2107  \\
50  &\hspace{1.2cm}  $3\%$   &\hspace{1.2cm} [-0.6196, 0.0796]    & \hspace{1.5cm}   0.2114  \\
50  &\hspace{1.2cm}  $5\%$   &\hspace{1.2cm} [-0.6206, 0.0804]    & \hspace{1.5cm}   0.2127  \\
\hline
100 &\hspace{1.2cm}  $0\%$   &\hspace{1.2cm} [-0.5918, 0.0588]    & \hspace{1.5cm}   0.1777  \\
100 &\hspace{1.2cm}  $3\%$   &\hspace{1.2cm} [-0.5927, 0.0595]    & \hspace{1.5cm}   0.1789  \\
100 &\hspace{1.2cm}  $5\%$   &\hspace{1.2cm} [-0.5943, 0.0607]    & \hspace{1.5cm}   0.1810  \\
\hline
300  &\hspace{1.2cm}  $0\%$   &\hspace{1.2cm} [-0.5616, 0.0359]    & \hspace{1.5cm}   0.1355  \\
300  &\hspace{1.2cm}  $3\%$   &\hspace{1.2cm} [-0.5634, 0.0372]    & \hspace{1.5cm}   0.1381  \\
300  &\hspace{1.2cm}  $5\%$   &\hspace{1.2cm} [-0.5663, 0.0394]    & \hspace{1.5cm}   0.1425  \\
\hline
500 &\hspace{1.2cm}  $0\%$   &\hspace{1.2cm} [-0.5518, 0.0284]    & \hspace{1.5cm}   0.1193 \\
500 &\hspace{1.2cm}  $3\%$   &\hspace{1.2cm} [-0.5540, 0.0301]    & \hspace{1.5cm}   0.1232 \\
500 &\hspace{1.2cm}  $5\%$   &\hspace{1.2cm} [-0.5577, 0.0329]    & \hspace{1.5cm}   0.1293 \\
\hline
1000 &\hspace{1.2cm}  $0\%$   &\hspace{1.2cm} [-0.5415, 0.0204]    & \hspace{1.5cm}   0.1005 \\
1000 &\hspace{1.2cm}  $3\%$   &\hspace{1.2cm} [-0.5447, 0.0229]    & \hspace{1.5cm}   0.1067 \\
1000 &\hspace{1.2cm}  $5\%$   &\hspace{1.2cm} [-0.5496, 0.0267]    & \hspace{1.5cm}   0.1156 \\
\hline\hline
\end{tabular}
\end{center}
\end{table}

\begin{table}[!ht]
\caption{Sensitivity on the $\hat a_V$ magnetic moment and the $\hat a_A$ electric dipole moment for the process
$e^-p \to e^-\gamma^*\gamma^*p \to e^-t\bar t p$ for $\sqrt{s}=7.07\hspace{0.8mm}TeV$ and ${\cal L}=50, 100, 300, 500,
1000\hspace{0.8mm}fb^{-1}$ at $90\%$ C.L..}
\begin{center}
 \begin{tabular}{cccc}
\hline\hline
\multicolumn{4}{c}{$\sqrt{s}=7.07\hspace{0.8mm}TeV$,  \hspace{1cm}  $90\%$ C.L.}\\
 \hline
 \cline{1-4} ${\cal L}\hspace{0.8mm}(fb^{-1})$  & \hspace{1.5cm} $\delta_{sys}$ & \hspace{1.5cm} $\hat a_V$   & $\hspace{1.7cm} |\hat a_A|$ \\
\hline
50  &\hspace{1.2cm}  $0\%$   &\hspace{1.2cm} [-0.6434, 0.0972]    & \hspace{1.5cm}   0.2107  \\
50  &\hspace{1.2cm}  $3\%$   &\hspace{1.2cm} [-0.6441, 0.0978]    & \hspace{1.5cm}   0.2114  \\
50  &\hspace{1.2cm}  $5\%$   &\hspace{1.2cm} [-0.6454, 0.0987]    & \hspace{1.5cm}   0.2127  \\
\hline
100 &\hspace{1.2cm}  $0\%$   &\hspace{1.2cm} [-0.6106, 0.0729]    & \hspace{1.5cm}   0.2009  \\
100 &\hspace{1.2cm}  $3\%$   &\hspace{1.2cm} [-0.6117, 0.0737]    & \hspace{1.5cm}   0.2022  \\
100 &\hspace{1.2cm}  $5\%$   &\hspace{1.2cm} [-0.6137, 0.0752]    & \hspace{1.5cm}   0.2045  \\
\hline
300  &\hspace{1.2cm}  $0\%$   &\hspace{1.2cm} [-0.5736, 0.0451]    & \hspace{1.5cm}   0.1533  \\
300  &\hspace{1.2cm}  $3\%$   &\hspace{1.2cm} [-0.5758, 0.0467]    & \hspace{1.5cm}   0.1563  \\
300  &\hspace{1.2cm}  $5\%$   &\hspace{1.2cm} [-0.5793, 0.0494]    & \hspace{1.5cm}   0.1612  \\
\hline
500 &\hspace{1.2cm}  $0\%$   &\hspace{1.2cm} [-0.5614, 0.0357]    & \hspace{1.5cm}   0.1351 \\
500 &\hspace{1.2cm}  $3\%$   &\hspace{1.2cm} [-0.5642, 0.0379]    & \hspace{1.5cm}   0.1394 \\
500 &\hspace{1.2cm}  $5\%$   &\hspace{1.2cm} [-0.5688, 0.0413]    & \hspace{1.5cm}   0.1463 \\
\hline
1000 &\hspace{1.2cm}  $0\%$   &\hspace{1.2cm} [-0.5485, 0.0259]    & \hspace{1.5cm}   0.1137 \\
1000 &\hspace{1.2cm}  $3\%$   &\hspace{1.2cm} [-0.5526, 0.0290]    & \hspace{1.5cm}   0.1207 \\
1000 &\hspace{1.2cm}  $5\%$   &\hspace{1.2cm} [-0.5587, 0.0336]    & \hspace{1.5cm}   0.1308 \\
\hline\hline
\end{tabular}
\end{center}
\end{table}

\begin{table}[!ht]
\caption{Sensitivity on the $\hat a_V$ magnetic moment and the $\hat a_A$ electric dipole moment for the process
$e^-p \to e^-\gamma^*\gamma^*p \to e^-t\bar t p$ for $\sqrt{s}=7.07\hspace{0.8mm}TeV$ and ${\cal L}=50, 100, 300, 500,
1000\hspace{0.8mm}fb^{-1}$ at $95\%$ C.L..}
\begin{center}
 \begin{tabular}{cccc}
\hline\hline
\multicolumn{4}{c}{$\sqrt{s}=7.07\hspace{0.8mm}TeV$,  \hspace{1cm}  $95\%$ C.L.}\\
 \hline
 \cline{1-4} ${\cal L}\hspace{0.8mm}(fb^{-1})$  & \hspace{1.5cm} $\delta_{sys}$ & \hspace{1.5cm} $\hat a_V$   & $\hspace{1.7cm} |\hat a_A|$ \\
\hline
50   &\hspace{1.2cm} $0\%$   &\hspace{1.2cm} [-0.6959, 0.1357]    & \hspace{1.5cm}   0.2922  \\
50  &\hspace{1.2cm}  $3\%$   &\hspace{1.2cm} [-0.6969, 0.1364]    & \hspace{1.5cm}   0.2931  \\
50  &\hspace{1.2cm}  $5\%$   &\hspace{1.2cm} [-0.6986, 0.1377]    & \hspace{1.5cm}   0.2948  \\
\hline
100 &\hspace{1.2cm}  $0\%$   &\hspace{1.2cm} [-0.6518, 0.1035]    & \hspace{1.5cm}   0.2471  \\
100 &\hspace{1.2cm}  $3\%$   &\hspace{1.2cm} [-0.6534, 0.1046]    & \hspace{1.5cm}   0.2487  \\
100 &\hspace{1.2cm}  $5\%$   &\hspace{1.2cm} [-0.6560, 0.1065]    & \hspace{1.5cm}   0.2515  \\
\hline
300  &\hspace{1.2cm}  $0\%$   &\hspace{1.2cm} [-0.6007, 0.0654]    & \hspace{1.5cm}   0.1889  \\
300  &\hspace{1.2cm}  $3\%$   &\hspace{1.2cm} [-0.6037, 0.0677]    & \hspace{1.5cm}   0.1926  \\
300  &\hspace{1.2cm}  $5\%$   &\hspace{1.2cm} [-0.6087, 0.0714]    & \hspace{1.5cm}   0.1986  \\
\hline
500 &\hspace{1.2cm}  $0\%$   &\hspace{1.2cm} [-0.5833, 0.0523]    & \hspace{1.5cm}   0.1665 \\
500 &\hspace{1.2cm}  $3\%$   &\hspace{1.2cm} [-0.5873, 0.0554]    & \hspace{1.5cm}   0.1719 \\
500 &\hspace{1.2cm}  $5\%$   &\hspace{1.2cm} [-0.5938, 0.0603]    & \hspace{1.5cm}   0.1803 \\
\hline
1000 &\hspace{1.2cm}  $0\%$   &\hspace{1.2cm} [-0.5648, 0.0383]    & \hspace{1.5cm}   0.1403 \\
1000 &\hspace{1.2cm}  $3\%$   &\hspace{1.2cm} [-0.5706, 0.0427]    & \hspace{1.5cm}   0.1489 \\
1000 &\hspace{1.2cm}  $5\%$   &\hspace{1.2cm} [-0.5794, 0.0494]    & \hspace{1.5cm}   0.1613 \\
\hline\hline
\end{tabular}
\end{center}
\end{table}

\begin{table}[!ht]
\caption{Sensitivity on the $\hat a_V$ magnetic moment and the $\hat a_A$ electric dipole moment for the process
$e^-p \to e^-\gamma^*\gamma^*p \to e^-t\bar t p$ for $\sqrt{s}=10\hspace{0.8mm}TeV$ and ${\cal L}=50, 100, 300, 500,
1000\hspace{0.8mm}fb^{-1}$ at $68\%$ C.L..}
\begin{center}
 \begin{tabular}{cccc}
\hline\hline
\multicolumn{4}{c}{$\sqrt{s}=10\hspace{0.8mm}TeV$,  \hspace{1cm}  $68\%$ C.L.}\\
 \hline
 \cline{1-4} ${\cal L}\hspace{0.8mm}(fb^{-1})$  & \hspace{1.5cm} $\delta_{sys}$ & \hspace{1.5cm} $\hat a_V$   & $\hspace{1.7cm} |\hat a_A|$ \\
\hline
50  &\hspace{1.2cm}  $0\%$   &\hspace{1.2cm} [-0.5292, 0.0689]    & \hspace{1.5cm}   0.1852  \\
50  &\hspace{1.2cm}  $3\%$   &\hspace{1.2cm} [-0.5298, 0.0694]    & \hspace{1.5cm}   0.1860  \\
50  &\hspace{1.2cm}  $5\%$   &\hspace{1.2cm} [-0.5310, 0.0703]    & \hspace{1.5cm}   0.1875  \\
\hline
100 &\hspace{1.2cm}  $0\%$   &\hspace{1.2cm} [-0.5077, 0.0511]    & \hspace{1.5cm}   0.1562  \\
100 &\hspace{1.2cm}  $3\%$   &\hspace{1.2cm} [-0.5087, 0.0519]    & \hspace{1.5cm}   0.1576  \\
100 &\hspace{1.2cm}  $5\%$   &\hspace{1.2cm} [-0.5104, 0.0533]    & \hspace{1.5cm}   0.1601  \\
\hline
300 &\hspace{1.2cm}  $0\%$   &\hspace{1.2cm} [-0.4839, 0.0311]    & \hspace{1.5cm}   0.1191  \\
300 &\hspace{1.2cm}  $3\%$   &\hspace{1.2cm} [-0.4857, 0.0327]    & \hspace{1.5cm}   0.1222  \\
300 &\hspace{1.2cm}  $5\%$   &\hspace{1.2cm} [-0.4888, 0.0352]    & \hspace{1.5cm}   0.1273  \\
\hline
500 &\hspace{1.2cm}  $0\%$   &\hspace{1.2cm} [-0.4761, 0.0245]    & \hspace{1.5cm}   0.1049 \\
500 &\hspace{1.2cm}  $3\%$   &\hspace{1.2cm} [-0.4785, 0.0266]    & \hspace{1.5cm}   0.1094 \\
500 &\hspace{1.2cm}  $5\%$   &\hspace{1.2cm} [-0.4823, 0.0298]    & \hspace{1.5cm}   0.1163 \\
\hline
1000 &\hspace{1.2cm} $0\%$   &\hspace{1.2cm} [-0.4680, 0.0177]    & \hspace{1.5cm}   0.0883 \\
1000 &\hspace{1.2cm} $3\%$   &\hspace{1.2cm} [-0.4714, 0.0205]    & \hspace{1.5cm}   0.0955 \\
1000 &\hspace{1.2cm} $5\%$   &\hspace{1.2cm} [-0.4763, 0.0247]    & \hspace{1.5cm}   0.1053 \\
\hline\hline
\end{tabular}
\end{center}
\end{table}

\begin{table}[!ht]
\caption{Sensitivity on the $\hat a_V$ magnetic moment and the $\hat a_A$ electric dipole moment for the process
$e^-p \to e^-\gamma^*\gamma^*p \to e^-t\bar t p$ for $\sqrt{s}=10\hspace{0.8mm}TeV$ and ${\cal L}=50, 100, 300, 500,
1000\hspace{0.8mm}fb^{-1}$ at $90\%$ C.L..}
\begin{center}
 \begin{tabular}{cccc}
\hline\hline
\multicolumn{4}{c}{$\sqrt{s}=10\hspace{0.8mm}TeV$,  \hspace{1cm}  $90\%$ C.L.}\\
 \hline
 \cline{1-4} ${\cal L}\hspace{0.8mm}(fb^{-1})$  & \hspace{1.5cm} $\delta_{sys}$ & \hspace{1.5cm} $\hat a_V$   & $\hspace{1.7cm} |\hat a_A|$ \\
\hline
50  &\hspace{1.2cm}  $0\%$   &\hspace{1.2cm} [-0.5485, 0.0848]    & \hspace{1.5cm}   0.2092  \\
50  &\hspace{1.2cm}  $3\%$   &\hspace{1.2cm} [-0.5492, 0.0854]    & \hspace{1.5cm}   0.2101  \\
50  &\hspace{1.2cm}  $5\%$   &\hspace{1.2cm} [-0.5506, 0.0865]    & \hspace{1.5cm}   0.2117  \\
\hline
100 &\hspace{1.2cm}  $0\%$   &\hspace{1.2cm} [-0.5226, 0.0634]    & \hspace{1.5cm}   0.1766  \\
100 &\hspace{1.2cm}  $3\%$   &\hspace{1.2cm} [-0.5238, 0.0644]    & \hspace{1.5cm}   0.1782  \\
100 &\hspace{1.2cm}  $5\%$   &\hspace{1.2cm} [-0.5259, 0.0661]    & \hspace{1.5cm}   0.1809  \\
\hline
300 &\hspace{1.2cm}  $0\%$   &\hspace{1.2cm} [-0.4934, 0.0391]    & \hspace{1.5cm}   0.1347  \\
300 &\hspace{1.2cm}  $3\%$   &\hspace{1.2cm} [-0.4957, 0.0410]    & \hspace{1.5cm}   0.1383  \\
300 &\hspace{1.2cm}  $5\%$   &\hspace{1.2cm} [-0.4994, 0.0441]    & \hspace{1.5cm}   0.1440  \\
\hline
500 &\hspace{1.2cm}  $0\%$   &\hspace{1.2cm} [-0.4837, 0.0309]    & \hspace{1.5cm}   0.1187 \\
500 &\hspace{1.2cm}  $3\%$   &\hspace{1.2cm} [-0.4867, 0.0334]    & \hspace{1.5cm}   0.1238 \\
500 &\hspace{1.2cm}  $5\%$   &\hspace{1.2cm} [-0.4914, 0.0374]    & \hspace{1.5cm}   0.1316 \\
\hline
1000 &\hspace{1.2cm} $0\%$   &\hspace{1.2cm} [-0.4735, 0.0224]    & \hspace{1.5cm}   0.1000 \\
1000 &\hspace{1.2cm} $3\%$   &\hspace{1.2cm} [-0.4778, 0.0260]    & \hspace{1.5cm}   0.1081 \\
1000 &\hspace{1.2cm} $5\%$   &\hspace{1.2cm} [-0.4840, 0.0311]    & \hspace{1.5cm}   0.1192 \\
\hline\hline
\end{tabular}
\end{center}
\end{table}

\begin{table}[!ht]
\caption{Sensitivity on the $\hat a_V$ magnetic moment and the $\hat a_A$ electric dipole moment for the process
$e^-p \to e^-\gamma^*\gamma^*p \to e^-t\bar t p$ for $\sqrt{s}=10\hspace{0.8mm}TeV$ and ${\cal L}=50, 100, 300, 500,
1000\hspace{0.8mm}fb^{-1}$ at $95\%$ C.L..}
\begin{center}
 \begin{tabular}{cccc}
\hline\hline
\multicolumn{4}{c}{$\sqrt{s}=10\hspace{0.8mm}TeV$,  \hspace{1cm}  $95\%$ C.L.}\\
 \hline
 \cline{1-4} ${\cal L}\hspace{0.8mm}(fb^{-1})$  & \hspace{1.5cm} $\delta_{sys}$ & \hspace{1.5cm} $\hat a_V$   & $\hspace{1.7cm} |\hat a_A|$ \\
\hline
50  &\hspace{1.2cm}  $0\%$   &\hspace{1.2cm} [-0.5900, 0.1188]    & \hspace{1.5cm}   0.2567  \\
50  &\hspace{1.2cm}  $3\%$   &\hspace{1.2cm} [-0.5910, 0.1196]    & \hspace{1.5cm}   0.2578  \\
50  &\hspace{1.2cm}  $5\%$   &\hspace{1.2cm} [-0.5928, 0.1211]    & \hspace{1.5cm}   0.2598  \\
\hline
100 &\hspace{1.2cm}  $0\%$   &\hspace{1.2cm} [-0.5551, 0.0903]    & \hspace{1.5cm}   0.2171  \\
100 &\hspace{1.2cm}  $3\%$   &\hspace{1.2cm} [-0.5568, 0.0916]    & \hspace{1.5cm}   0.2191  \\
100 &\hspace{1.2cm}  $5\%$   &\hspace{1.2cm} [-0.5596, 0.0939]    & \hspace{1.5cm}   0.2224  \\
\hline
300 &\hspace{1.2cm}  $0\%$   &\hspace{1.2cm} [-0.5147, 0.0569]    & \hspace{1.5cm}   0.1660  \\
300 &\hspace{1.2cm}  $3\%$   &\hspace{1.2cm} [-0.5179, 0.0595]    & \hspace{1.5cm}   0.1704  \\
300 &\hspace{1.2cm}  $5\%$   &\hspace{1.2cm} [-0.5232, 0.0639]    & \hspace{1.5cm}   0.1774  \\
\hline
500 &\hspace{1.2cm}  $0\%$   &\hspace{1.2cm} [-0.5010, 0.0454]    & \hspace{1.5cm}   0.1464 \\
500 &\hspace{1.2cm}  $3\%$   &\hspace{1.2cm} [-0.5053, 0.0490]    & \hspace{1.5cm}   0.1527 \\
500 &\hspace{1.2cm}  $5\%$   &\hspace{1.2cm} [-0.5120, 0.0546]    & \hspace{1.5cm}   0.1622 \\
\hline
1000 &\hspace{1.2cm} $0\%$   &\hspace{1.2cm} [-0.4864, 0.0332]    & \hspace{1.5cm}   0.1233 \\
1000 &\hspace{1.2cm} $3\%$   &\hspace{1.2cm} [-0.4925, 0.0383]    & \hspace{1.5cm}   0.1334 \\
1000 &\hspace{1.2cm} $5\%$   &\hspace{1.2cm} [-0.5014, 0.0457]    & \hspace{1.5cm}   0.1470 \\
\hline\hline
\end{tabular}
\end{center}
\end{table}

\vspace{0.3cm}

\begin{center}
{\bf Acknowledgments}
\end{center}

A. G. R. and M. A. H. R. acknowledge support from SNI and PROFOCIE (M\'exico).

\pagebreak

\begin{figure}[t]
\centerline{\scalebox{0.8}{\includegraphics{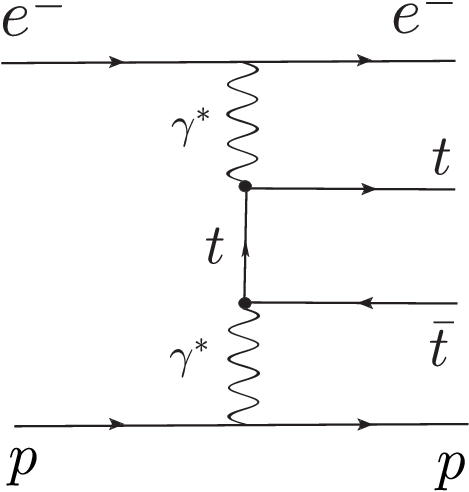}}}
\caption{ \label{fig:gamma1} A schematic diagram for the process
$e^-p \to e^-\gamma^*\gamma^*p \to e^-t\bar t p$.}
\label{Fig.1}
\end{figure}

\begin{figure}[t]
\centerline{\scalebox{0.8}{\includegraphics{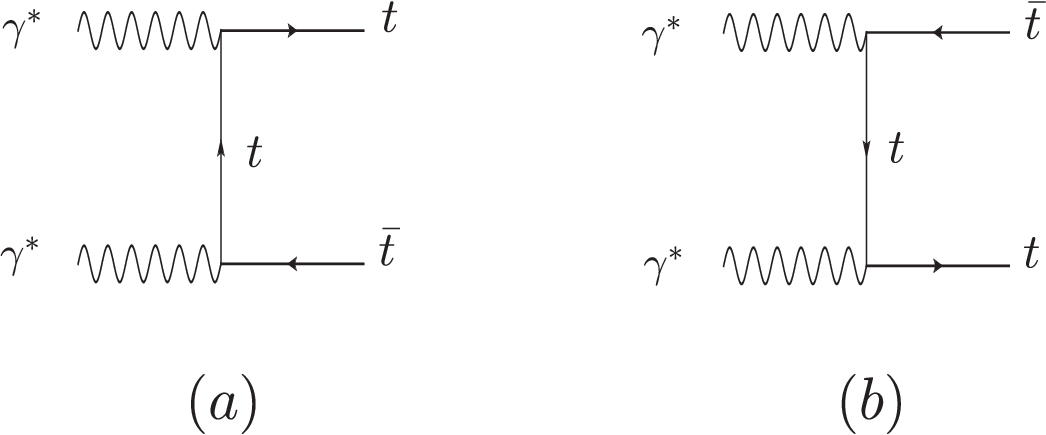}}}
\caption{ \label{fig:gamma2} Feynman diagrams contributing to the subprocess
$\gamma^*\gamma^* \to t \bar t$.}
\label{Fig.2}
\end{figure}

\begin{figure}[t]
\centerline{\scalebox{1.2}{\includegraphics{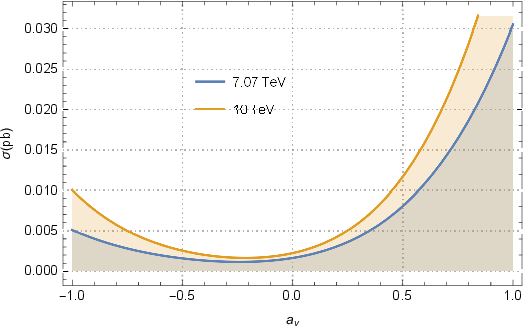}}}
\caption{ The total cross sections of the process
$e^-p \to e^-\gamma^*\gamma^*p \to e^-t\bar t p$ as a function of $\hat a_V$
for center-of-mass energies of $\sqrt{s}=7.07, 10$\hspace{0.8mm}$TeV$ at the FCC-he.}
\label{Fig.3}
\end{figure}

\begin{figure}[t]
\centerline{\scalebox{1.2}{\includegraphics{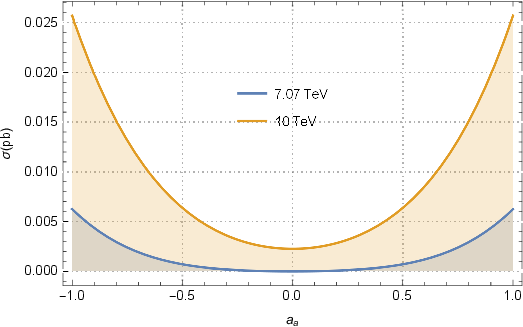}}}
\caption{ Same as in Fig. 3, but for $\hat a_A$.}
\label{Fig.4}
\end{figure}

\begin{figure}[t]
\centerline{\scalebox{1.2}{\includegraphics{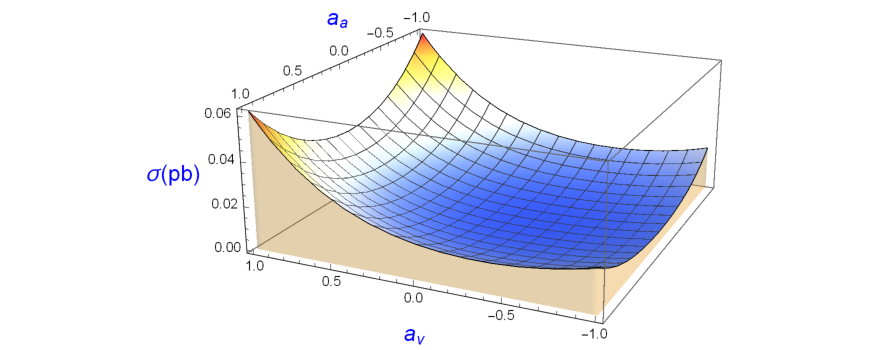}}}
\caption{ \label{fig:gamma1} The total cross sections of the process
$e^-p \to e^-\gamma^*\gamma^*p \to e^-t\bar t p$ as a function of $\hat a_V$ and $\hat a_A$
for center-of-mass energy of $\sqrt{s}=7.07$\hspace{0.8mm}$TeV$ at the FCC-he.}
\label{Fig.5}
\end{figure}

\begin{figure}[t]
\centerline{\scalebox{1.2}{\includegraphics{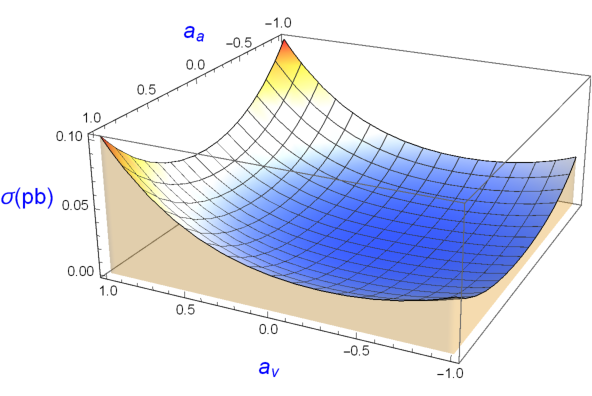}}}
\caption{ \label{fig:gamma2} Same as in Fig. 5, but for center-of-mass energy of
$\sqrt{s}=10$\hspace{0.8mm}$TeV$.}
\label{Fig.6}
\end{figure}

\begin{figure}[t]
\centerline{\scalebox{1.1}{\includegraphics{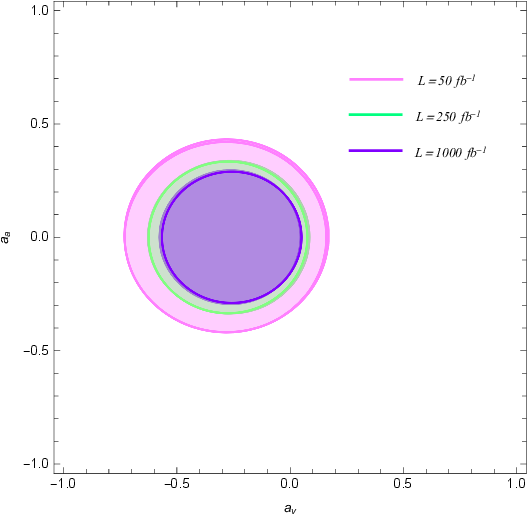}}}
\caption{ \label{fig:gamma1x} Sensitivity contours at the $95\% \hspace{1mm}C.L.$ in the
$\hat a_V-\hat a_A$ plane through the process $e^-p \to e^-\gamma^*\gamma^*p \to e^-t\bar t p$
for $\sqrt{s}=7.07$\hspace{0.8mm}$TeV$ at the FCC-he.}
\label{Fig.7}
\end{figure}

\begin{figure}[t]
\centerline{\scalebox{1.1}{\includegraphics{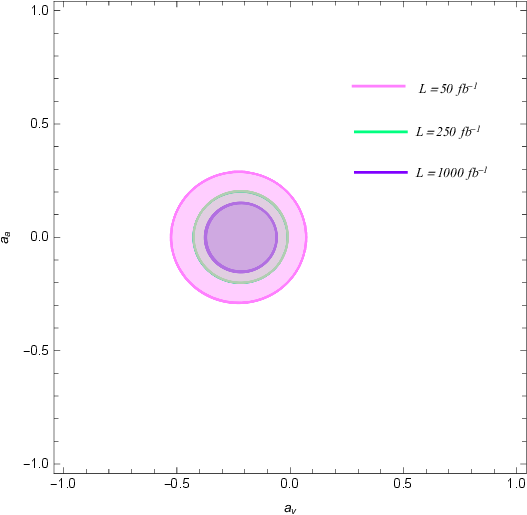}}}
\caption{ \label{fig:gamma2x} Same as in Fig. 7, but for $\sqrt{s}=10$\hspace{0.8mm}$TeV$.}
\label{Fig.8}
\end{figure}

\begin{figure}[t]
\centerline{\scalebox{1.2}{\includegraphics{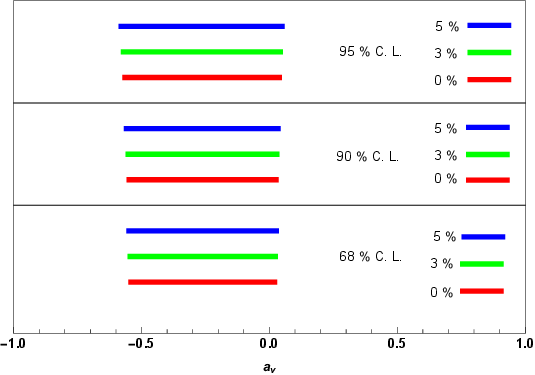}}}
\caption{ \label{fig:gamma15} The expected sensitivities of FCC-he measurements at
$\sqrt{s}=7.07\hspace{0.8mm}TeV$ to $\hat a_V$ in the process
$e^-p \to e^-\gamma^*\gamma^*p \to e^-t\bar t p$.
We assume systematic uncertainties $\delta_{sys}=0, 3, 5\hspace{0.8mm}\%$ and $68\%, 90\%, 95\%\hspace{0.8mm}C.L.$.}
\label{Fig.6}
\end{figure}

\begin{figure}[t]
\centerline{\scalebox{1.2}{\includegraphics{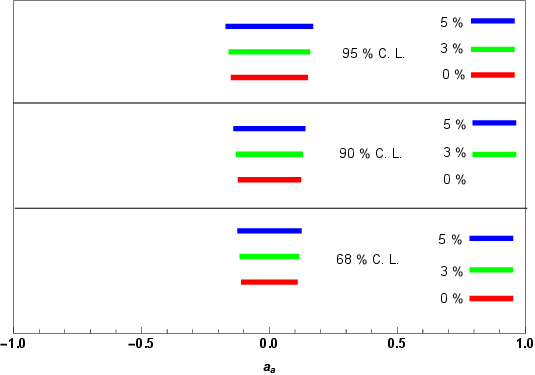}}}
\caption{ \label{fig:gamma6} Same as in Fig. 9, but for $\hat a_A$.}
\label{Fig.7}
\end{figure}

\begin{figure}[t]
\centerline{\scalebox{1.2}{\includegraphics{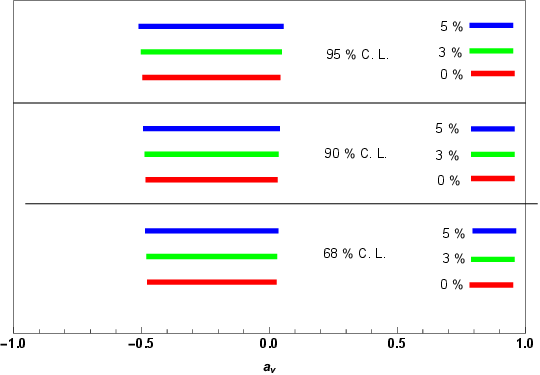}}}
\caption{ \label{fig:gamma6x} Same as in Fig. 9, but for $\sqrt{s}=10\hspace{0.8mm}TeV$.}
\label{Fig.8}
\end{figure}

\begin{figure}[t]
\centerline{\scalebox{1.2}{\includegraphics{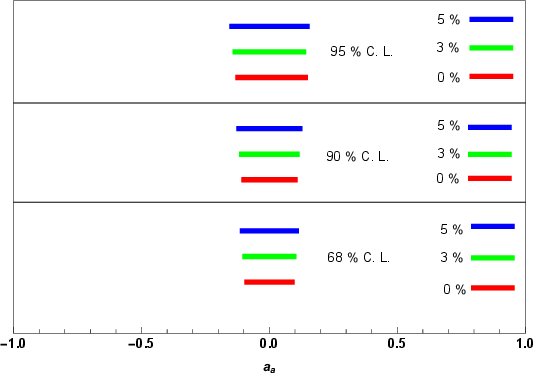}}}
\caption{ \label{fig:gamma6x} Same as in Fig. 11, but for $\hat a_A$.}
\label{Fig.8}
\end{figure}

\end{document}